\documentclass[aps,reprint,prl,twocolumn,showpacs,preprintnumbers,amsmath,
amssymb,nofootinbib,superscriptaddress,showkeys]{revtex4-1}

\usepackage{epsfig}
\usepackage{color}
\usepackage{slashed}

\begin{document}

\title{Heavy Antiquark--Diquark Symmetry and Heavy Hadron Molecules:\\
Are There Triply Heavy Pentaquarks?}

\author{Feng-Kun Guo}
\email{fkguo@hiskp.uni-bonn.de}
\affiliation{Helmholtz-Institut f\"ur Strahlen- und
             Kernphysik and Bethe Center for Theoretical Physics, \\
             Universit\"at Bonn,  D-53115 Bonn, Germany}
\author{Carlos Hidalgo-Duque}
\email{carloshd@ific.uv.es}
\affiliation{Instituto de F\'isica Corpuscular (IFIC),
             Centro Mixto CSIC-Universidad de Valencia,
             Institutos de Investigaci\'on de Paterna,
             Aptd. 22085, E-46071 Valencia, Spain}
\author{Juan Nieves}
\email{jmnieves@ific.uv.es}
\affiliation{Instituto de F\'isica Corpuscular (IFIC),
             Centro Mixto CSIC-Universidad de Valencia,
             Institutos de Investigaci\'on de Paterna,
             Aptd. 22085, E-46071 Valencia, Spain}
\author{Manuel Pav\'on Valderrama}
\email{pavonvalderrama@ipno.in2p3.fr}
\affiliation{Institut de Physique Nucl\'eaire,
             Universit\'e Paris-Sud, IN2P3/CNRS,
             F-91406 Orsay Cedex, France}

\begin{abstract}
\rule{0ex}{3ex}
We explore the consequences of heavy flavour, heavy quark spin
and heavy antiquark-diquark symmetries for hadronic molecules
within an effective field theory framework.
Owing to heavy antiquark-diquark symmetry, the doubly heavy baryons
have approximately the same light-quark
structure as the heavy antimesons.
As a consequence, the existence of a heavy meson-antimeson molecule
implies the possibility of a partner
composed of a heavy meson and a doubly-heavy baryon.
In this regard, the $D\bar D^*$ molecular nature of the $X(3872)$
will hint at the existence of several baryonic partners with isospin
$I=0$ and $J^P = {\frac{5}{2}}^-$ or $\frac32^-$.
Moreover, if the $Z_b(10650)$ turns out to be a $B^*\bar{B}^*$ bound state,
we can be confident of the existence of $\Xi_{bb}^* \bar{B}^*$
hadronic molecules with quantum numbers $I(J^P) = 1({\frac{1}{2}}^-)$
and $I(J^P) = 1({\frac{3}{2}}^-)$.
These states are of special interest since they can be
considered to be triply-heavy pentaquarks.
\end{abstract}


\maketitle

The spectroscopic properties of bound states tell us a great deal about
the symmetries and underlying dynamics of their components.
For instance, the hydrogen atom has been an extraordinary source of
information about several aspects of quantum electrodynamics,
from the accidental SO(4) symmetry in the spectrum
to the vacuum polarization, radiative corrections
and renormalization that are necessary
to explain the Lamb shift.
The classification of hadrons according to isospin, SU(3) flavour and so on
reveals the underlying strong dynamics binding
the hadrons and has been instrumental in the past
for the development of quantum chromodynamics (QCD).
Conversely a deeper understanding of QCD and its symmetries --- such as
heavy quark symmetries --- will eventually give new insights into the hadron 
spectrum.

Heavy hadron molecules are a type of exotic hadron
theorized more than thirty years ago~\cite{Voloshin:1976ap,DeRujula:1976qd}.
Their main component is a pair of heavy hadrons instead
of a quark--antiquark pair. 
The experimental advances in heavy quarkonium spectroscopy 
have identified several molecular candidates among the recently observed $XYZ$ 
states.
The most promising ones are the $X(3872)$~\cite{Choi:2003ue} and
the twin $Z_b(10610)$ and $Z_b(10650)$ states, to be called $Z_b$ and $Z_b'$,
respectively~\cite{Belle:2011aa,Adachi:2012im}.
The $Y(4260)$~\cite{Aubert:2005rm} might have finally revealed
its (so far conjectural~\cite{Ding:2008gr,Li:2013bca})
molecular nature~\cite{Wang:2013cya}
by being an intermediate step in the production of
the recent $Z_c(3900)$~\cite{BESIII:2013,Liu:2013xoa}
(which may also be molecular~\cite{Guo:2013sya,Wang:2013cya,Wilbring:2013cha}).
There are always competing explanations such as conventional heavy
quarkonia, tetraquarks or hybrid states which have the same quantum numbers.
It is thus challenging to distinguish the hadronic molecules from other
possibilities.
In this work, we will explore a model-independent approach which leads
to unique predictions for hadronic molecules.
A set of triply-heavy pentaquark-like molecules can be predicted as partners of
heavy mesonic molecules as a consequence of heavy quark symmetries.

Heavy hadronic molecules are very
interesting objects because they have an exceptionally high degrees of symmetry
stemming from their combined light and
heavy quark content~\cite{Guo:2009id,Bondar:2011ev,Voloshin:2011qa,Mehen:2011yh,
Valderrama:2012jv,Nieves:2012tt,HidalgoDuque:2012pq,Guo:2013sya}.
While the presence of heavy quarks imply that molecular states
are subject to heavy quark symmetries,
the light quarks allow to classify the molecular
states in isospin and SU(3) flavor multiplets~\cite{HidalgoDuque:2012pq}.
Among the manifestations of heavy quark symmetries, we can count
heavy quark spin symmetry (HQSS),
which implies that molecular states may appear in HQSS multiplets,
for instance (but not limited to) the $D_{s0}^*(2317)$ and
$D_{s1}(2460)$~\cite{Guo:2009id}, the $Z_b$ and
$Z_b'$~\cite{Bondar:2011ev,Voloshin:2011qa, Mehen:2011yh,Nieves:2012tt}.
From heavy flavor symmetry (HFS), we know that the interaction among
heavy hadrons is roughly independent on whether they contain
a charm or a bottom quark.
In this regard, the recently discovered $Z_c(3900)$ could very well be a
heavy flavour partner of the $Z_b$~\cite{Guo:2013sya}.
Last, there is a less explored type of heavy quark symmetry that is going
to have particularly interesting consequences:
heavy antiquark-diquark symmetry (HADS)~\cite{Savage:1990di}.

HADS states that
the two heavy quarks within a doubly heavy baryon behave approximately as a
heavy antiquark.
The heavy diquark component of the baryon forms a color anti-triplet
with a characteristic length scale of $1 / (m_Q v)$,
where $m_Q$ is the mass of the heavy quarks
and $v$ their velocity.
The length scale of the diquark is smaller than the typical QCD
length scale $1/\Lambda_{\rm QCD}$ and hence we can treat
the diquark as point-like if the quarks are heavy enough.
The consequence is that the light-quark cloud surrounding the heavy diquark
in a heavy baryon would be similar to the one around the heavy antiquark
in a heavy antimeson.
We expect violations of the order
of $\Lambda_{\rm QCD} / (m_Q v)$, instead of $\Lambda_{\rm QCD} / m_Q$
as in HQSS and HFS.
This translates into a $30-40\%$ uncertainty in the charm sector
and $15-20\%$ in the bottom one.
Yet even with this limitation HADS can be useful
in suggesting the possibility of new charmed molecules (see also
Ref.~\cite{Cohen:2006jg} for a discussion of this symmetry
in the charm sector),
while for bottom ones concrete predictions can be made, as we will
show in what follows.

We consider a state molecular if its most important component is
the set of hadrons conforming the molecule, where the other components
such as compact multiquarks play a minor role in its description.
In general this is only true if the separation among the hadrons is big enough
as for them to retain their individual character.
This suggests that genuine molecular states show a clear separation
of scales between their long and short range structure and are
thus amenable to an effective field theory (EFT)
treatment~\cite{Braaten:2003he,Fleming:2007rp,Mehen:2011yh,Valderrama:2012jv}
(analogous to nuclear EFT,
see Refs.~\cite{Epelbaum:2008ga,Machleidt:2011zz} for reviews).
Among the theoretical advantages of EFT a very interesting
one is power counting:
we can expand any physical quantity as a power series in terms of
the expansion parameter $(Q / M)$, where $Q$ is a typical low energy
scale (for instance, the inverse of the size of the molecular state)
and $M$ the high energy scale at which the EFT description stops
being valid (the inverse size of the hadrons).
The heavy hadrons are nonrelativistic:
thus we can define a hadron-hadron potential that admits
the low energy expansion $V_{HH}^{\rm EFT} = V_{HH}^{\rm LO} + V_{HH}^{\rm NLO} 
+ \dots$, where ${\rm LO}$ stands for ``leading order'', ${\rm NLO}$
for ``next-to-leading order'' and so on.
In addition, the interaction among the heavy hadrons forming a molecule
is nonperturbative so that we have to iterate the ${\rm LO}$
(i.e. the most important) piece of the EFT potential.
Due to the exploratory character of this work,
we will not go beyond ${\rm LO}$.

Now we will consider the case of heavy meson-antimeson molecules
with the light and heavy quarks being $q = u,d$ and $Q = c,b$.
The EFT potential is subjected to the constraints of chiral symmetry
and heavy quark symmetries.
At ${\rm LO}$ there are only contact range interactions, with pion exchanges
entering as a correction at ${\rm NLO}$,
for more details, see~\cite{Valderrama:2012jv} \footnote{An exception are 
isoscalar bottom molecules, for which the strength of the OPE potential is  
considerable and
hence should be included at ${\rm LO}$ unless
the molecular state is very shallow.}.
The application of HQSS leads to the following ${\rm LO}$
potential~\cite{Nieves:2012tt,HidalgoDuque:2012pq}.
\begin{eqnarray}
V^{\rm LO}_{P \bar P}(\vec{q}\, , {0^{++}})
&=& C_{Ia} \label{eq:coni} \, , \\
V^{\rm LO}_{P^* {\bar P} / P {\bar P}^*}(\vec{q}\, ,{1^{+-}})
&=& V^{\rm LO}_{P^* {\bar P^*}}(\vec{q}\, ,{1^{+-}}) = C_{Ia} - C_{Ib} \,
,
\\
V^{\rm LO}_{P^* {\bar P}  / P {\bar P}^*}(\vec{q}\, ,{1^{++}})
&=& V^{\rm LO}_{P^* {\bar P}^* }(\vec{q}\, ,{2^{++}}) = C_{Ia} + C_{Ib} \,
,
\\
V^{\rm LO}_{P^* {\bar P}^* }(\vec{q}\, ,{0^{++}})
&=& C_{Ia} - 2\,C_{Ib} \, ,
\label{eq:conf}
\end{eqnarray}
where the subscript indicates the particle channel ($P = {D, \bar{B}}$
and $P^* = {D^*, \bar{B}^*}$), ${\vec q}$ is the momentum exchanged
by the heavy meson and antimeson and $J^{PC}$ indicates the quantum
numbers.
The subscript $I = 0,1$ indicates the isospin of the molecule (unless stated
otherwise, we are working in the isospin symmetric limit).
For each isospin, the ${\rm LO}$ potential depends on two parameters,
$C_{Ia}$ and $C_{Ib}$, that determine the mass of up to six states.

In turn, HFS implies that the previous potential does not depend
on the flavor of heavy quarks contained in the mesons~\cite{Guo:2013sya}.
As a consequence of this symmetry we can expect a particular pattern
of states in the charm sector to repeat itself in the bottom one,
though the binding energies will be different.

Finally, we consider the interaction between doubly-heavy baryons
$\Xi_{Q_1Q_2},\Xi_{Q_1Q_2}^{*}$ ($Q_{1,2}=c,b$,  total spin of the heavy pair
$s_{Q_1Q_2}=1$ and $J^P= \frac12^+$ and $\frac32^+$, respectively) or the
$J^P=\frac12^+$ $\Xi_{bc}'$ ($s_{bc}=0$), and a heavy
meson $P^{(*)}$. HADS allows us to write the LO
$\Xi_{Q_1Q_2}^{(*)} P^{(*)}$ potential in terms of the same
counter-terms $C_{Ia}$ and $C_{Ib}$ that appear in the LO
meson-antimeson potential.
The analysis of the light quark components
in the heavy baryon-meson system leads to potentials listed in
the Table~\ref{tab:potentials}, 
\begin{table*}[tb]
\caption{\label{tab:potentials} LO potentials and quantum numbers for various
doubly-heavy baryon--heavy meson systems.
}
\begin{ruledtabular}
\begin{tabular}{l | c c c c c c c c c c}
       States & $\Xi_{Q_1Q_2}P$ & $\Xi_{Q_1Q_2}P^*$ &
       $\Xi_{Q_1Q_2}P^*$  & $\Xi_{Q_1Q_2 }^*P$
       & $\Xi_{Q_1Q_2}^*P^*$ & $\Xi_{Q_1Q_2}^*P^*$ & $\Xi_{Q_1Q_2}^*P^*$ & 
$\Xi_{bc}'P$ &
       $\Xi_{bc}'P^*$ &       $\Xi_{bc}'P^*$ \\
       \hline
       $J^{P}$ & $\frac12^-$ & $\frac12^-$ & $\frac32^-$  & $\frac32^-$ &
       $\frac12^-$ & $\frac32^-$ & $\frac52^-$ & $\frac12^-$ & $\frac12^-$ &
       $\frac32^-$ \\
       $V^{\rm LO}$ & $C_{Ia}$ & $C_{Ia}+\frac23 C_{Ib}$ &
       $C_{Ia}-\frac13 C_{Ib}$ & $C_{Ia}$ & $C_{Ia}-\frac53 C_{Ib}$ &
       $C_{Ia}-\frac23 C_{Ib}$ & $C_{Ia}+C_{Ib}$ & $C_{Ia}$ & $C_{Ia}-2C_{Ib}$ &
       $C_{Ia}+C_{Ib}$ \\
   \end{tabular}
\end{ruledtabular}
\end{table*}
%
from which we can derive the spectrum of the heavy baryon-meson molecules.

To estimate the binding energies of the molecules we solve the
Lippmann-Schwinger equation and look for the poles of the $T$-matrix.
The EFT potential is singular when iterated: we have to regularize
and renormalize the potential to make predictions.
For the renormalization we employ a gaussian regulator with the cut-offs
$\Lambda = 0.5\,{\rm GeV}$ and $1\,{\rm GeV}$, and the couplings
$C_{Ia}$ and $C_{Ib}$ will depend on $\Lambda$.
The complete procedure and the choice of the cut-off window is explained
in detail in Refs.~\cite{Nieves:2012tt,HidalgoDuque:2012pq}.
For the meson masses, we take isospin averaged values $M_D = 1867.24$~MeV, 
$M_{D^*} =
2008.63$~MeV, $M_B = 5279.34$~MeV, $M_{B^*} = 5325.1$~MeV and $M_X = 3871.68
\,{\rm MeV}$~\cite{PDG}.
The mass of the $Z_b^{(\prime)}$ reported in
Ref.~\cite{Belle:2011aa} (\cite{Adachi:2012cx}) is
$1\sigma$ higher (overlaps)  with the corresponding $B^{(*)}\bar B^*$ threshold.
However, the location of the $Z_b$'s may depend on the parametrization
employed for them, as shown in Ref.~\cite{Cleven:2011gp}.
Therefore, we simply assume that the binding energy of the $Z_b$ is
$2.0 \pm 2.0\,{\rm MeV}$, as in Ref.~\cite{Guo:2013sya}.
The doubly charmed baryons were  only reported by the SELEX
Collaboration~\cite{Mattson:2002vu,Moinester:2002uw, Ocherashvili:2004hi}.
However, the measured masses are lower than expectations in most of the model
and lattice calculations, and the observed isospin splittings seem too large to
be accommodated in QCD~\cite{Brodsky:2011zs}.
Thus, we will use a recent lattice calculation for the masses, $M_{\Xi_{cc}} =
3606 \pm 22$~MeV and $M_{\Xi_{cc}^*} = 3706 \pm 28$~MeV~\cite{Namekawa:2013vu}.
For the doubly bottom baryons, there is no experimental
observation yet, and the lattice QCD predictions are $M_{\Xi_{bb}} = 10127 \pm
13 {}^{+12}_{-26}$~MeV and $M_{\Xi_{bb}^*} = 10151 \pm 14 
{}^{+16}_{-25}$~MeV~\cite{Lewis:2008fu}.
Their ventral values will be used.
Furthermore, we take constituent quark model predictions for the $\Xi_{bc}'$ and 
$\Xi^*_{bc}$ masses,
6958 and 6996~MeV, respectively~\cite{Albertus:2009ww}.
Predictions will be made for the binding energies instead of masses to avoid
introducing the lattice QCD errors of the baryon masses into the results.
Finally, the HQSS/HFS uncertainty in the counter-terms is assumed to be
$20\%$($7\%$) in the charm (bottom sector), while for HADS we use $40\%$
($20\%$).
They are assumed to be uncorrelated.
We will also use a $30\%$ HADS uncertainty for the $\Xi_{bc}'$
systems.
We will not show them explicitly when writing down the value
of the counter-terms, yet we will take them into account.

\begin{table*}[ttt]
\caption{\label{tab:predictions}
Predictions of the doubly-heavy baryon--heavy meson
molecules.
The isoscalar states are related to the $X(3872)$, and the error in
their binding energies is a consequence of the approximate nature of HADS.
The isovector states are determined by the $Z_b(10610,10650)$ and the
isovector component of the $X$.
In this part, different error sources have been taken into account: the
uncertainty in the $Z_b$ binding, in the isospin breaking decays of the $X$
and in the HADS breaking. For simplicity, we only show an unique error obtained
by adding in quadratures all the previous ones.
Here, $M_{\rm th}$ represents the threshold, and all masses are given in units
of MeV.  When we decrease the
strength of the potential to account for the various uncertainties, in some 
cases (marked with $\dagger$ in the table)  the bound state pole reaches the 
threshold and the state becomes virtual. The cases with a virtual state pole 
at the central value are marked by [V], for which $\dagger\dagger$ 
means that the pole 
evolves into a bound state one and N/A means that the pole is far from the 
threshold with a momentum larger than 1~GeV so that it is both undetectable and 
beyond the EFT range.}
\begin{ruledtabular}
\begin{tabular}{l c c c c c}
       State & $I(J^{P})$ & $V^{\rm LO}$ & Thresholds &
       Mass ($\Lambda=0.5$ GeV) & Mass ($\Lambda=1$ GeV) \\ \hline
       $\Xi_{cc}^* D^*$ & $0({\frac{5}{2}}^-)$ & $C_{0a}+C_{0b}$ & $5715$
       & $\left( M_{\rm th} - 10\right)^{+10}_{-15}$ & $\left(M_{\rm th} - 
19\right)^{\dagger}_{-44}$ \\
       $\Xi_{cc}^* \bar{B}^*$ & $0({\frac{5}{2}}^-)$ & $C_{0a}+C_{0b}$ &
       $9031$ & $\left(M_{\rm th} - 21\right)^{+16}_{-19}$ & $\left(M_{\rm th} - 
53\right)^{+45}_{-59}$ \\
       $\Xi_{bb}^* D^*$ & $0({\frac{5}{2}}^-)$ & $C_{0a}+C_{0b}$ & $12160$
       & $\left(M_{\rm th} - 15\right)^{+9}_{-11}$ & $\left(M_{\rm th} - 
35\right)^{+25}_{-31}$ \\
       $\Xi_{bb}^* \bar{B}^*$ & $0({\frac{5}{2}}^-)$
       & $C_{0a}+C_{0b}$ & $15476$
       & $\left(M_{\rm th} - 29\right)^{+12}_{-13}$ & $\left(M_{\rm th} - 
83\right)^{+38}_{-40}$ \\
       $\Xi_{bc}' D^*$ & $0({\frac{3}{2}}^-)$ & $C_{0a}+C_{0b}$ & $8967$
       & $\left(M_{\rm th} - 14\right)^{+11}_{-13}$ & $\left(M_{\rm th} - 
30\right)^{+27}_{-40}$ \\
       $\Xi_{bc}' \bar{B}^*$ & $0({\frac{3}{2}}^-)$
       & $C_{0a}+C_{0b}$ & $12283$
       & $\left(M_{\rm th} - 27\right)^{+15}_{-16}$ & $\left(M_{\rm th} - 
74\right)^{+45}_{-51}$ \\
$\Xi_{bc}^* D^*$ & $0({\frac{5}{2}}^-)$ & $C_{0a}+C_{0b}$ & $9005$
& $\left(M_{\rm th} - 14\right)^{+11}_{-13}$ & $\left(M_{\rm th} -
30\right)^{+27}_{-40}$ \\
$\Xi_{bc}^* \bar{B}^*$ & $0({\frac{5}{2}}^-)$ & $C_{0a}+C_{0b}$ & $12321$
& $\left(M_{\rm th} -27\right)^{+15}_{-16}$ & $\left(M_{\rm th} 
-74\right)^{+46}_{-51}$ \\\hline
       $\Xi_{bb}\bar{B}$ & $1({\frac{1}{2}}^-)$ & $C_{1a}$ & $15406$
       & $\left(M_{\rm th} - 0.3\right)_{-2.5}^{\dagger}$ &
       $\left(M_{\rm th} -  12\right)^{+11}_{-15}$ \\
       $\Xi_{bb}\bar{B}^*$ & $1({\frac{1}{2}}^-)$ &
       $C_{1a}+\frac{2}{3}\,C_{1b}$ & $15452$
       & ($M_{\rm th}-0.9$)[V]$_{\dagger\dagger}^\text{N/A}$
       & $\left(M_{\rm th} - 16\right)^{+14}_{-17}$ \\
       $\Xi_{bb}\bar{B}^*$ & $1({\frac{3}{2}}^-)$
       & $C_{1a}-\frac{1}{3}\,C_{1b}$ & $15452$
       & $\left(M_{\rm th} - 1.2\right)^{\dagger}_{-2.9}$
       & $\left(M_{\rm th} - 10\right)^{+9}_{-13}$ \\
       $\Xi_{bb}^*\bar{B}$ & $1({\frac{3}{2}}^-)$ & $C_{1a}$ & $15430$
       & $\left(M_{\rm th} - 0.3\right)^{\dagger}_{-2.4}$
       & $\left(M_{\rm th} - 12\right)^{+11}_{-13}$ \\
       $\Xi_{bb}^*\bar{B}^*$ & $1({\frac{1}{2}}^-)$
       & $C_{1a}-\frac{5}{3}\,C_{1b}$ & $15476$
       & $\left(M_{\rm th} - 8\right)^{+8}_{-7}$
       & $\left(M_{\rm th} - 5\right)^{\dagger}_{-8}$ \\
       $\Xi_{bb}^*\bar{B}^*$ & $1({\frac{3}{2}}^-)$ & 
$C_{1a}-\frac{2}{3}\,C_{1b}$ & $15476$
       & $\left(M_{\rm th} - 2.5\right)^{\dagger}_{-3.6}$
       & $\left(M_{\rm th} -9\right)^{+9}_{-11}$ \\
       $\Xi_{bb}^*\bar{B}^*$ & $1({\frac{5}{2}}^-)$ & $C_{1a}+C_{1b}$ & $15476$
       & $\left(M_{\rm th}-4.3\right)$[V]$_{+3.3}^\text{N/A}$
       & $\left(M_{\rm th} - 18\right)^{+17}_{-19}$ \\
   \end{tabular}
\end{ruledtabular}
\end{table*}

We begin by considering the $X(3872)$ as a pure isoscalar $1^{++}$
$D\bar{D^*}$ molecule as in \cite{Nieves:2012tt}.
The ${\rm LO}$ potential is given by the counter-term combination
$C_{0X} \equiv C_{0a} + C_{0b}$, which is identical to the one
appearing in the family
of $\Xi_{Q_1Q_2}^* P^*$ with $J = \frac{5}{2}^-$  and the $\Xi_{bc}'P^*$ systems
with $J^P=\frac32^-$.
We have $C_{0X} = -1.94\,(-0.79)\,{\rm fm}^2$
for $\Lambda = 0.5\,(1)\,{\rm GeV}$ \cite{Nieves:2012tt}.
Bound state solutions are found in all the considered systems, though the
$\Xi^*_{cc} D^*$ system can be very loosely bound due to the large uncertainty
of the LO potential, and the predictions can be found in
Table~\ref{tab:predictions}.
In addition, it is more than probable that the isoscalar $\Xi^{(*\prime)}_{bc} 
{B}^*$ and $\Xi^*_{bb} \bar{B}^*$ molecules require nonperturbative OPE owing
to their heavy reduced mass. Though, the nonperturbative OPE will  modify the
binding energies, we expect, however, that these systems will remain still bound.

Now we continue with what can be deduced from the $Z_b^{(\prime)}$
as isovector $1^{+-}$ $B^{(*)} \bar{B}^*$ molecular states.
As can be seen from Table~\ref{tab:potentials}, there is no exact match among
the LO potential for the $Z_b$'s, $C_{1Z} \equiv C_{1a} - C_{1b}$, and the six 
possible $\Xi_{Q_1Q_2}^{(*)}
P^{(*)}$ configurations.
Yet, the $\frac{3}{2}^-$
$\Xi_{Q_1Q_2}^{*} P^{*}$ configuration has coupling:
$C_{1a} - \frac{2}{3}\,C_{1b} = C_{1Z} + \frac{1}{3}\,C_{1b}$.
As far as the relative contribution of the $C_{1b}$ coupling
is not excessive, a hadronic molecule, either as a bound or virtual state, looks 
probable.
Other two interesting configurations are the $\frac{1}{2}^-$
$\Xi_{Q_1Q_2}^{*} P^{*}$ and $\frac{3}{2}^-$ $\Xi_{Q_1Q_2} P^{*}$
systems, for which the couplings read
$C_{1Z} \mp \frac{2}{3}\,C_{1b}$.
Depending on the sign and size of $C_{1b}$ at least one of
the two configurations should bind.

All this indicates that the isospin-1 doubly-heavy baryon--meson molecules are
probable, but a further assessment requires the determination of both
$C_{1a}$ and $C_{1b}$.
From the $Z_b$'s we obtain~\cite{Guo:2013sya}
$C_{1Z} = 
-\left(0.75^{+0.15}_{-0.28}\right)\,[-\left(0.30^{+0.03}_{-0.07}\right)]\,{\rm 
fm}^2$
for $\Lambda = 0.5\,{\rm GeV}\, [1\,{\rm GeV}]$, where the errors come
from the uncertainties in the binding energy.
But for disentangling the $C_{1a}$ and $C_{1b}$ couplings a second
source of information is necessary.
For that we will use the isospin symmetry breaking of the $X(3872)$,
which offers a window into the interaction in the isovector
$1^{++}$ $D\bar{D}^*$ channel~\cite{HidalgoDuque:2012pq}.
The decay of the $X(3872)$ into the isovector $J/\psi 2\pi$ channel
indicates that the $X$ is not a pure isoscalar state,
but contains a small isovector component.
The branching ratio of the isovector $J/\psi 2\pi$ to the isoscalar
$J/\psi 3\pi$ decays constrains the size of this component and
hence the strength of the interaction in the isovector
channel~\cite{HidalgoDuque:2012pq}.
We find $C_{1X} = -(0.13 \pm 0.40)\,[-(0.39\pm 0.09)]\,{\rm fm}^2$
for $\Lambda = 0.5\,{\rm GeV}\, [1\,{\rm GeV}]$, where the errors reflect the
experimental uncertainty in the branching ratio.\footnote{The central value of 
$C_{1X}$ differs from that quoted in
\cite{HidalgoDuque:2012pq} by an amount that is  around 10\% of its error. This 
is because of the use of different values  for the $X$ resonance mass.}
Using the formulas $C_{1a} =
(C_{1X}+C_{1Z})/2$ and $C_{1b} = (C_{1X}-C_{1Z})/2$, we obtain
$C_{1a} = -(0.44\pm 0.24) \,\,[-(0.35\pm 0.06)]\,{\rm fm}^2$ and $C_{1b} =
(0.31\pm 0.24)\,\,[-(0.05\pm 0.06)]\,{\rm fm}^2$ (the errors shown are for
guidance only and have been obtained by adding in quadratures those
quoted for $C_{1X}$ and $C_{1Z}$).
We see that $C_{1b}$ is either positive or, if negative,
extremely small and that $|C_{1b}| < |C_{1a}|$, which
already contains a lot of information about
the possible molecular states.
We show the predictions in Table~\ref{tab:predictions}, where
the uncertainties in the binding energies come from the errors
in $C_{1X}$ and $C_{1Z}$, the additional HQSS/HFS $20\%$ error
(as part of the information comes from the charm sector)
and from the expected $20\%$ violation of HADS.

In the isovector sector, all configurations are plausible molecular candidates.
However, when we take into account the various uncertainties of
the current approach, we cannot discard in all cases
the appearance of virtual states instead
of proper bound molecules.
The most promising predictions are the $\frac{1}{2}^-$ and
$\frac{3}{2}^-$ $\Xi_{bb}^{*}\bar{B}^{*}$  molecules,
for which binding is moderately robust against
the different error sources.

To confirm these states from the theoretical side we need to pinpoint
the value of $C_{1b}$ more accurately.
This could be done either by more accurate measurements of the $X$ isospin
breaking ratio or, better yet, by the eventual discovery of HQSS partners
of the $Z_b$'s, the $W_b$ states proposed in Ref.~\cite{Voloshin:2011qa}.
Notice that all the isospin-1 triply-heavy molecules are very interesting
in the sense that they have a non-trivial pentaquark component.
We point out that though heavy pentaquarks have been predicted in the
literature on the basis of
several arguments~\cite{Genovese:1997tm,Stewart:2004pd,Cohen:2005bx},
this is the first prediction of a triply heavy one.

To summarize, we have studied the implications of HADS
(plus HQSS and HFS) for heavy hadronic molecules.
As a consequence of this symmetry, we can be confident about the existence
of doubly-heavy baryon--heavy meson (and eventually doubly-heavy
baryon--antibaryon: $\Xi_{Q_1Q_2}^{(*)}$--$\Xi_{\bar Q_1 \bar Q_2}^{(*)}$)
partners of heavy meson--antimeson molecules. From the assumption that the
$X(3872)$ and the $Z_b(10610/10650)$ are molecular states we can predict the
existence of the exotic pentaquark-like partners of these states.
We notice that phase space forbids any of the predicted molecules to
decay through the strong decays of their components. One of the possible strong
decay channels is a triply-heavy baryon plus one or more pions. Such a decay 
involves exchanging a heavy quark and a light
quark so that it would have a small partial width. The $\Xi_{Q_1Q_2}P$ in a $D$
wave could be the dominant decay channel of the $\Xi_{Q_1Q_2}^*P^*$ states with
$J^P=\frac32^-$ and $\frac52^-$. However, if the binding energy is so small that 
the binding
momentum is much smaller than the pion mass, the predicted state should be
quite stable.
It would be intriguing if any of the predicted states can be found in
high-energy hadron colliders and heavy ion collisions.

\begin{acknowledgments}
We would like to thank Ulf-G. Mei{\ss}ner for a careful reading of the
manuscript. FKG acknowledges the Theory Division of IHEP in Beijing, where part
of the work was done, for the hospitality. This work is supported in part by the 
DFG
and the NSFC through funds provided to the Sino-German CRC 110 ``Symmetries and
the Emergence of Structure in QCD'', by the NSFC (Grant No. 11165005), by the
Spanish Ministerio de Econom\'\i a y Competitividad and European FEDER funds
under the contract  FIS2011-28853-C02-02 and the Spanish
Consolider-Ingenio 2010 Programme CPAN (CSD2007-00042),  by Generalitat
Valenciana under contract PROMETEO/2009/0090  and by the EU
HadronPhysics2 project, grant no. 227431.
\end{acknowledgments}

\bibliography{hads}

\begin{thebibliography}{10}%
\makeatletter
\providecommand \@ifxundefined [1]{%
 \ifx #1\undefined \expandafter \@firstoftwo
 \else \expandafter \@secondoftwo
\fi
}%
\providecommand \@ifnum [1]{%
 \ifnum #1\expandafter \@firstoftwo
 \else \expandafter \@secondoftwo
\fi
}%
\providecommand \enquote [1]{``#1''}%
\providecommand \bibnamefont  [1]{#1}%
\providecommand \bibfnamefont [1]{#1}%
\providecommand \citenamefont [1]{#1}%
\providecommand\href[0]{\@sanitize\@href}%
\providecommand\@href[1]{\endgroup\@@startlink{#1}\endgroup\@@href}%
\providecommand\@@href[1]{#1\@@endlink}%
\providecommand \@sanitize [0]{\begingroup\catcode`\&12\catcode`\#12\relax}%
\@ifxundefined \pdfoutput {\@firstoftwo}{%
 \@ifnum{\z@=\pdfoutput}{\@firstoftwo}{\@secondoftwo}%
}{%
 \providecommand\@@startlink[1]{\leavevmode\special{html:<a href="#1">}}%
 \providecommand\@@endlink[0]{\special{html:</a>}}%
}{%
 \providecommand\@@startlink[1]{%
  \leavevmode
  \pdfstartlink
   attr{/Border[0 0 1 ]/H/I/C[0 1 1]}%
   user{/Subtype/Link/A<</Type/Action/S/URI/URI(#1)>>}%
  \relax
 }%
 \providecommand\@@endlink[0]{\pdfendlink}%
}%
\providecommand \url  [0]{\begingroup\@sanitize \@url }%
\providecommand \@url [1]{\endgroup\@href {#1}{\urlprefix}}%
\providecommand \urlprefix [0]{URL }%
\providecommand \Eprint[0]{\href }%
\@ifxundefined \urlstyle {%
  \providecommand \doi [1]{doi:\discretionary{}{}{}#1}%
}{%
  \providecommand \doi [0]{doi:\discretionary{}{}{}\begingroup
  \urlstyle{rm}\Url }%
}%
\providecommand \doibase [0]{http://dx.doi.org/}%
\providecommand \Doi[1]{\href{\doibase#1}}%
\providecommand \bibAnnote [3]{%
  \BibitemShut{#1}%
  \begin{quotation}\noindent
    \textsc{Key:}\ #2\\\textsc{Annotation:}\ #3%
  \end{quotation}%
}%
\providecommand \bibAnnoteFile [2]{%
  \IfFileExists{#2}{\bibAnnote {#1} {#2} {\input{#2}}}{}%
}%
\providecommand \typeout [0]{\immediate \write \m@ne }%
\providecommand \selectlanguage [0]{\@gobble}%
\providecommand \bibinfo [0]{\@secondoftwo}%
\providecommand \bibfield [0]{\@secondoftwo}%
\providecommand \translation [1]{[#1]}%
\providecommand \BibitemOpen[0]{}%
\providecommand \bibitemStop [0]{}%
\providecommand \bibitemNoStop [0]{.\EOS\space}%
\providecommand \EOS [0]{\spacefactor3000\relax}%
\providecommand \BibitemShut [1]{\csname bibitem#1\endcsname}%
\bibitem{Voloshin:1976ap}%
  \BibitemOpen
  \bibfield{author}{%
  \bibinfo {author} {\bibfnamefont{M.}~\bibnamefont{Voloshin}}\ and\ \bibinfo
  {author} {\bibfnamefont{L.}~\bibnamefont{Okun}},\ }%
  \bibfield{journal}{%
  \bibinfo {journal} {JETP Lett.}\ }%
  \textbf{\bibinfo {volume} {23}},\ \bibinfo {pages} {333} (\bibinfo {year}
  {1976})%
  \bibAnnoteFile{NoStop}{Voloshin:1976ap}%
\bibitem{DeRujula:1976qd}%
  \BibitemOpen
  \bibfield{author}{%
  \bibinfo {author} {\bibfnamefont{A.}~\bibnamefont{De~Rujula}}, \bibinfo
  {author} {\bibfnamefont{H.}~\bibnamefont{Georgi}},\ and\ \bibinfo {author}
  {\bibfnamefont{S.}~\bibnamefont{Glashow}},\ }%
  \bibfield{journal}{%
  \Doi{10.1103/PhysRevLett.38.317}{\bibinfo {journal} {Phys.Rev.Lett.}}\ }%
  \textbf{\bibinfo {volume} {38}},\ \bibinfo {pages} {317} (\bibinfo {year}
  {1977})%
  \bibAnnoteFile{NoStop}{DeRujula:1976qd}%
\bibitem{Choi:2003ue}%
  \BibitemOpen
  \bibfield{author}{%
  \bibinfo {author} {\bibfnamefont{S.~K.}\ \bibnamefont{Choi}} \emph{et~al.}
  (\bibinfo {collaboration} {Belle}),\ }%
  \bibfield{journal}{%
  \Doi{10.1103/PhysRevLett.91.262001}{\bibinfo {journal} {Phys. Rev. Lett.}}\
  }%
  \textbf{\bibinfo {volume} {91}},\ \bibinfo {pages} {262001} (\bibinfo {year}
  {2003}),\ \Eprint{http://arxiv.org/abs/hep-ex/0309032}{arXiv:hep-ex/0309032}%
  \bibAnnoteFile{NoStop}{Choi:2003ue}%
\bibitem{Belle:2011aa}%
  \BibitemOpen
  \bibfield{author}{%
  \bibinfo {author} {\bibfnamefont{A.}~\bibnamefont{Bondar}} \emph{et~al.}
  (\bibinfo {collaboration} {Belle Collaboration}),\ }%
  \bibfield{journal}{%
  \Doi{10.1103/PhysRevLett.108.122001}{\bibinfo {journal} {Phys.Rev.Lett.}}\ }%
  \textbf{\bibinfo {volume} {108}},\ \bibinfo {pages} {122001} (\bibinfo {year}
  {2012}),\ \Eprint{http://arxiv.org/abs/1110.2251}{arXiv:1110.2251 [hep-ex]}%
  \bibAnnoteFile{NoStop}{Belle:2011aa}%
\bibitem{Adachi:2012im}%
  \BibitemOpen
  \bibfield{author}{%
  \bibinfo {author} {\bibfnamefont{I.}~\bibnamefont{Adachi}} \emph{et~al.}
  (\bibinfo {collaboration} {Belle Collaboration})}%
   (\bibinfo {year} {2012}),\
  \Eprint{http://arxiv.org/abs/1207.4345}{arXiv:1207.4345 [hep-ex]}%
  \bibAnnoteFile{NoStop}{Adachi:2012im}%
\bibitem{Aubert:2005rm}%
  \BibitemOpen
  \bibfield{author}{%
  \bibinfo {author} {\bibfnamefont{B.}~\bibnamefont{Aubert}} \emph{et~al.}
  (\bibinfo {collaboration} {BABAR Collaboration}),\ }%
  \bibfield{journal}{%
  \Doi{10.1103/PhysRevLett.95.142001}{\bibinfo {journal} {Phys.Rev.Lett.}}\ }%
  \textbf{\bibinfo {volume} {95}},\ \bibinfo {pages} {142001} (\bibinfo {year}
  {2005}),\ \Eprint{http://arxiv.org/abs/hep-ex/0506081}{arXiv:hep-ex/0506081
  [hep-ex]}%
  \bibAnnoteFile{NoStop}{Aubert:2005rm}%
\bibitem{Ding:2008gr}%
  \BibitemOpen
  \bibfield{author}{%
  \bibinfo {author} {\bibfnamefont{G.-J.}\ \bibnamefont{Ding}},\ }%
  \bibfield{journal}{%
  \Doi{10.1103/PhysRevD.79.014001}{\bibinfo {journal} {Phys.Rev.}}\ }%
  \textbf{\bibinfo {volume} {D79}},\ \bibinfo {pages} {014001} (\bibinfo {year}
  {2009}),\ \Eprint{http://arxiv.org/abs/0809.4818}{arXiv:0809.4818 [hep-ph]}%
  \bibAnnoteFile{NoStop}{Ding:2008gr}%
\bibitem{Li:2013bca}%
  \BibitemOpen
  \bibfield{author}{%
  \bibinfo {author} {\bibfnamefont{M.-T.}\ \bibnamefont{Li}}, \bibinfo {author}
  {\bibfnamefont{W.-L.}\ \bibnamefont{Wang}}, \bibinfo {author}
  {\bibfnamefont{Z.-Y.}\ \bibnamefont{Zhang}},\ and\ \bibinfo {author}
  {\bibfnamefont{Y.-B.}\ \bibnamefont{Dong}}}%
   (\bibinfo {year} {2013}),\
  \Eprint{http://arxiv.org/abs/1303.4140}{arXiv:1303.4140 [nucl-th]}%
  \bibAnnoteFile{NoStop}{Li:2013bca}%
\bibitem{Wang:2013cya}%
  \BibitemOpen
  \bibfield{author}{%
  \bibinfo {author} {\bibfnamefont{Q.}~\bibnamefont{Wang}}, \bibinfo {author}
  {\bibfnamefont{C.}~\bibnamefont{Hanhart}},\ and\ \bibinfo {author}
  {\bibfnamefont{Q.}~\bibnamefont{Zhao}}}%
   (\bibinfo {year} {2013}),\
  \Eprint{http://arxiv.org/abs/1303.6355}{arXiv:1303.6355 [hep-ph]}%
  \bibAnnoteFile{NoStop}{Wang:2013cya}%
\bibitem{BESIII:2013}%
  \BibitemOpen
  \bibfield{author}{%
  \bibinfo {author} {\bibfnamefont{M.}~\bibnamefont{Ablikim}} \emph{et~al.}
  (\bibinfo {collaboration} {BESIII Collaboration})}%
   (\bibinfo {year} {2013}),\
  \Eprint{http://arxiv.org/abs/1303.5949}{arXiv:1303.5949 [hep-ex]}%
  \bibAnnoteFile{NoStop}{BESIII:2013}%
\bibitem{Liu:2013xoa}%
  \BibitemOpen
  \bibfield{author}{%
  \bibinfo {author} {\bibfnamefont{Z.}~\bibnamefont{Liu}} \emph{et~al.}
  (\bibinfo {collaboration} {Belle Collaboration})}%
   (\bibinfo {year} {2013}),\
  \Eprint{http://arxiv.org/abs/1304.0121}{arXiv:1304.0121 [hep-ex]}%
  \bibAnnoteFile{NoStop}{Liu:2013xoa}%
\bibitem{Guo:2013sya}%
  \BibitemOpen
  \bibfield{author}{%
  \bibinfo {author} {\bibfnamefont{F.-K.}\ \bibnamefont{Guo}}, \bibinfo
  {author} {\bibfnamefont{C.}~\bibnamefont{Hidalgo-Duque}}, \bibinfo {author}
  {\bibfnamefont{J.}~\bibnamefont{Nieves}},\ and\ \bibinfo {author}
  {\bibfnamefont{M.}~\bibnamefont{Pavon~Valderrama}}}%
   (\bibinfo {year} {2013}),\
  \Eprint{http://arxiv.org/abs/1303.6608}{arXiv:1303.6608 [hep-ph]}%
  \bibAnnoteFile{NoStop}{Guo:2013sya}%
\bibitem{Wilbring:2013cha}%
  \BibitemOpen
  \bibfield{author}{%
  \bibinfo {author} {\bibfnamefont{E.}~\bibnamefont{Wilbring}}, \bibinfo
  {author} {\bibfnamefont{H.~W.}\ \bibnamefont{Hammer}},\ and\ \bibinfo
  {author} {\bibfnamefont{U.~G.}\ \bibnamefont{Mei{\ss}ner}}}%
   (\bibinfo {year} {2013}),\
  \Eprint{http://arxiv.org/abs/1304.2882}{arXiv:1304.2882 [hep-ph]}%
  \bibAnnoteFile{NoStop}{Wilbring:2013cha}%
\bibitem{Guo:2009id}%
  \BibitemOpen
  \bibfield{author}{%
  \bibinfo {author} {\bibfnamefont{F.-K.}\ \bibnamefont{Guo}}, \bibinfo
  {author} {\bibfnamefont{C.}~\bibnamefont{Hanhart}},\ and\ \bibinfo {author}
  {\bibfnamefont{U.-G.}\ \bibnamefont{Mei{\ss}ner}},\ }%
  \bibfield{journal}{%
  \Doi{10.1103/PhysRevLett.102.242004}{\bibinfo {journal} {Phys. Rev. Lett.}}\
  }%
  \textbf{\bibinfo {volume} {102}},\ \bibinfo {pages} {242004} (\bibinfo {year}
  {2009}),\ \Eprint{http://arxiv.org/abs/0904.3338}{arXiv:0904.3338 [hep-ph]}%
  \bibAnnoteFile{NoStop}{Guo:2009id}%
\bibitem{Bondar:2011ev}%
  \BibitemOpen
  \bibfield{author}{%
  \bibinfo {author} {\bibfnamefont{A.}~\bibnamefont{Bondar}}, \bibinfo {author}
  {\bibfnamefont{A.}~\bibnamefont{Garmash}}, \bibinfo {author}
  {\bibfnamefont{A.}~\bibnamefont{Milstein}}, \bibinfo {author}
  {\bibfnamefont{R.}~\bibnamefont{Mizuk}},\ and\ \bibinfo {author}
  {\bibfnamefont{M.}~\bibnamefont{Voloshin}},\ }%
  \bibfield{journal}{%
  \Doi{10.1103/PhysRevD.84.054010}{\bibinfo {journal} {Phys.Rev.}}\ }%
  \textbf{\bibinfo {volume} {D84}},\ \bibinfo {pages} {054010} (\bibinfo {year}
  {2011}),\ \Eprint{http://arxiv.org/abs/1105.4473}{arXiv:1105.4473 [hep-ph]}%
  \bibAnnoteFile{NoStop}{Bondar:2011ev}%
\bibitem{Voloshin:2011qa}%
  \BibitemOpen
  \bibfield{author}{%
  \bibinfo {author} {\bibfnamefont{M.}~\bibnamefont{Voloshin}},\ }%
  \bibfield{journal}{%
  \Doi{10.1103/PhysRevD.84.031502}{\bibinfo {journal} {Phys.Rev.}}\ }%
  \textbf{\bibinfo {volume} {D84}},\ \bibinfo {pages} {031502} (\bibinfo {year}
  {2011}),\ \Eprint{http://arxiv.org/abs/1105.5829}{arXiv:1105.5829 [hep-ph]}%
  \bibAnnoteFile{NoStop}{Voloshin:2011qa}%
\bibitem{Mehen:2011yh}%
  \BibitemOpen
  \bibfield{author}{%
  \bibinfo {author} {\bibfnamefont{T.}~\bibnamefont{Mehen}}\ and\ \bibinfo
  {author} {\bibfnamefont{J.~W.}\ \bibnamefont{Powell}},\ }%
  \bibfield{journal}{%
  \Doi{10.1103/PhysRevD.84.114013}{\bibinfo {journal} {Phys.Rev.}}\ }%
  \textbf{\bibinfo {volume} {D84}},\ \bibinfo {pages} {114013} (\bibinfo {year}
  {2011}),\ \Eprint{http://arxiv.org/abs/1109.3479}{arXiv:1109.3479 [hep-ph]}%
  \bibAnnoteFile{NoStop}{Mehen:2011yh}%
\bibitem{Valderrama:2012jv}%
  \BibitemOpen
  \bibfield{author}{%
  \bibinfo {author} {\bibfnamefont{M.}~\bibnamefont{Pavon~Valderrama}},\ }%
  \bibfield{journal}{%
  \Doi{10.1103/PhysRevD.85.114037}{\bibinfo {journal} {Phys.Rev.}}\ }%
  \textbf{\bibinfo {volume} {D85}},\ \bibinfo {pages} {114037} (\bibinfo {year}
  {2012}),\ \Eprint{http://arxiv.org/abs/1204.2400}{arXiv:1204.2400 [hep-ph]}%
  \bibAnnoteFile{NoStop}{Valderrama:2012jv}%
\bibitem{Nieves:2012tt}%
  \BibitemOpen
  \bibfield{author}{%
  \bibinfo {author} {\bibfnamefont{J.}~\bibnamefont{Nieves}}\ and\ \bibinfo
  {author} {\bibfnamefont{M.}~\bibnamefont{Pavon~Valderrama}},\ }%
  \bibfield{journal}{%
  \Doi{10.1103/PhysRevD.86.056004}{\bibinfo {journal} {Phys.Rev.}}\ }%
  \textbf{\bibinfo {volume} {D86}},\ \bibinfo {pages} {056004} (\bibinfo {year}
  {2012}),\ \Eprint{http://arxiv.org/abs/1204.2790}{arXiv:1204.2790 [hep-ph]}%
  \bibAnnoteFile{NoStop}{Nieves:2012tt}%
\bibitem{HidalgoDuque:2012pq}%
  \BibitemOpen
  \bibfield{author}{%
  \bibinfo {author} {\bibfnamefont{C.}~\bibnamefont{Hidalgo-Duque}}, \bibinfo
  {author} {\bibfnamefont{J.}~\bibnamefont{Nieves}},\ and\ \bibinfo {author}
  {\bibfnamefont{M.}~\bibnamefont{Pavon~Valderrama}},\ }%
  \bibfield{journal}{%
  \Doi{10.1103/PhysRevD.87.076006}{\bibinfo {journal} {Phys.Rev.}}\ }%
  \textbf{\bibinfo {volume} {D87}},\ \bibinfo {pages} {076006} (\bibinfo {year}
  {2012}),\ \Eprint{http://arxiv.org/abs/1210.5431}{arXiv:1210.5431 [hep-ph]}%
  \bibAnnoteFile{NoStop}{HidalgoDuque:2012pq}%
\bibitem{Savage:1990di}%
  \BibitemOpen
  \bibfield{author}{%
  \bibinfo {author} {\bibfnamefont{M.~J.}\ \bibnamefont{Savage}}\ and\ \bibinfo
  {author} {\bibfnamefont{M.~B.}\ \bibnamefont{Wise}},\ }%
  \bibfield{journal}{%
  \Doi{10.1016/0370-2693(90)90035-5}{\bibinfo {journal} {Phys.Lett.}}\ }%
  \textbf{\bibinfo {volume} {B248}},\ \bibinfo {pages} {177} (\bibinfo {year}
  {1990})%
  \bibAnnoteFile{NoStop}{Savage:1990di}%
\bibitem{Cohen:2006jg}%
  \BibitemOpen
  \bibfield{author}{%
  \bibinfo {author} {\bibfnamefont{T.~D.}\ \bibnamefont{Cohen}}\ and\ \bibinfo
  {author} {\bibfnamefont{P.~M.}\ \bibnamefont{Hohler}},\ }%
  \bibfield{journal}{%
  \Doi{10.1103/PhysRevD.74.094003}{\bibinfo {journal} {Phys.Rev.}}\ }%
  \textbf{\bibinfo {volume} {D74}},\ \bibinfo {pages} {094003} (\bibinfo {year}
  {2006}),\ \Eprint{http://arxiv.org/abs/hep-ph/0606084}{arXiv:hep-ph/0606084
  [hep-ph]}%
  \bibAnnoteFile{NoStop}{Cohen:2006jg}%
\bibitem{Braaten:2003he}%
  \BibitemOpen
  \bibfield{author}{%
  \bibinfo {author} {\bibfnamefont{E.}~\bibnamefont{Braaten}}\ and\ \bibinfo
  {author} {\bibfnamefont{M.}~\bibnamefont{Kusunoki}},\ }%
  \bibfield{journal}{%
  \Doi{10.1103/PhysRevD.69.074005}{\bibinfo {journal} {Phys.Rev.}}\ }%
  \textbf{\bibinfo {volume} {D69}},\ \bibinfo {pages} {074005} (\bibinfo {year}
  {2004}),\ \Eprint{http://arxiv.org/abs/hep-ph/0311147}{arXiv:hep-ph/0311147
  [hep-ph]}%
  \bibAnnoteFile{NoStop}{Braaten:2003he}%
\bibitem{Fleming:2007rp}%
  \BibitemOpen
  \bibfield{author}{%
  \bibinfo {author} {\bibfnamefont{S.}~\bibnamefont{Fleming}}, \bibinfo
  {author} {\bibfnamefont{M.}~\bibnamefont{Kusunoki}}, \bibinfo {author}
  {\bibfnamefont{T.}~\bibnamefont{Mehen}},\ and\ \bibinfo {author}
  {\bibfnamefont{U.}~\bibnamefont{van Kolck}},\ }%
  \bibfield{journal}{%
  \Doi{10.1103/PhysRevD.76.034006}{\bibinfo {journal} {Phys. Rev.}}\ }%
  \textbf{\bibinfo {volume} {D76}},\ \bibinfo {pages} {034006} (\bibinfo {year}
  {2007}),\ \Eprint{http://arxiv.org/abs/hep-ph/0703168}{arXiv:hep-ph/0703168}%
  \bibAnnoteFile{NoStop}{Fleming:2007rp}%
\bibitem{Epelbaum:2008ga}%
  \BibitemOpen
  \bibfield{author}{%
  \bibinfo {author} {\bibfnamefont{E.}~\bibnamefont{Epelbaum}}, \bibinfo
  {author} {\bibfnamefont{H.-W.}\ \bibnamefont{Hammer}},\ and\ \bibinfo
  {author} {\bibfnamefont{U.-G.}\ \bibnamefont{Mei{\ss}ner}},\ }%
  \bibfield{journal}{%
  \Doi{10.1103/RevModPhys.81.1773}{\bibinfo {journal} {Rev. Mod. Phys.}}\ }%
  \textbf{\bibinfo {volume} {81}},\ \bibinfo {pages} {1773} (\bibinfo {year}
  {2009}),\ \Eprint{http://arxiv.org/abs/0811.1338}{arXiv:0811.1338 [nucl-th]}%
  \bibAnnoteFile{NoStop}{Epelbaum:2008ga}%
\bibitem{Machleidt:2011zz}%
  \BibitemOpen
  \bibfield{author}{%
  \bibinfo {author} {\bibfnamefont{R.}~\bibnamefont{Machleidt}}\ and\ \bibinfo
  {author} {\bibfnamefont{D.}~\bibnamefont{Entem}},\ }%
  \bibfield{journal}{%
  \Doi{10.1016/j.physrep.2011.02.001}{\bibinfo {journal} {Phys.Rept.}}\ }%
  \textbf{\bibinfo {volume} {503}},\ \bibinfo {pages} {1} (\bibinfo {year}
  {2011}),\ \Eprint{http://arxiv.org/abs/1105.2919}{arXiv:1105.2919 [nucl-th]}%
  \bibAnnoteFile{NoStop}{Machleidt:2011zz}%
\bibitem{PDG}%
  \BibitemOpen
  \bibfield{author}{%
  \bibinfo {author} {\bibfnamefont{J.}~\bibnamefont{Beringer}} \emph{et~al.}
  (\bibinfo {collaboration} {Particle Data Group}),\ }%
  \bibfield{journal}{%
  \bibinfo {journal} {Phys. Rev. D}\ }%
  \textbf{\bibinfo {volume} {86}},\ \bibinfo {pages} {010001} (\bibinfo {year}
  {2012})%
  \bibAnnoteFile{NoStop}{PDG}%
\bibitem{Adachi:2012cx}%
  \BibitemOpen
  \bibfield{author}{%
  \bibinfo {author} {\bibfnamefont{I.}~\bibnamefont{Adachi}} \emph{et~al.}
  (\bibinfo {collaboration} {Belle Collaboration})}%
   (\bibinfo {year} {2012}),\
  \Eprint{http://arxiv.org/abs/1209.6450}{arXiv:1209.6450 [hep-ex]}%
  \bibAnnoteFile{NoStop}{Adachi:2012cx}%
\bibitem{Cleven:2011gp}%
  \BibitemOpen
  \bibfield{author}{%
  \bibinfo {author} {\bibfnamefont{M.}~\bibnamefont{Cleven}}, \bibinfo {author}
  {\bibfnamefont{F.-K.}\ \bibnamefont{Guo}}, \bibinfo {author}
  {\bibfnamefont{C.}~\bibnamefont{Hanhart}},\ and\ \bibinfo {author}
  {\bibfnamefont{U.-G.}\ \bibnamefont{Mei{\ss}ner}},\ }%
  \bibfield{journal}{%
  \Doi{10.1140/epja/i2011-11120-6}{\bibinfo {journal} {Eur.Phys.J.}}\ }%
  \textbf{\bibinfo {volume} {A47}},\ \bibinfo {pages} {120} (\bibinfo {year}
  {2011}),\ \Eprint{http://arxiv.org/abs/1107.0254}{arXiv:1107.0254 [hep-ph]}%
  \bibAnnoteFile{NoStop}{Cleven:2011gp}%
\bibitem{Mattson:2002vu}%
  \BibitemOpen
  \bibfield{author}{%
  \bibinfo {author} {\bibfnamefont{M.}~\bibnamefont{Mattson}} \emph{et~al.}
  (\bibinfo {collaboration} {SELEX Collaboration}),\ }%
  \bibfield{journal}{%
  \Doi{10.1103/PhysRevLett.89.112001}{\bibinfo {journal} {Phys.Rev.Lett.}}\ }%
  \textbf{\bibinfo {volume} {89}},\ \bibinfo {pages} {112001} (\bibinfo {year}
  {2002}),\ \Eprint{http://arxiv.org/abs/hep-ex/0208014}{arXiv:hep-ex/0208014
  [hep-ex]}%
  \bibAnnoteFile{NoStop}{Mattson:2002vu}%
\bibitem{Moinester:2002uw}%
  \BibitemOpen
  \bibfield{author}{%
  \bibinfo {author} {\bibfnamefont{M.}~\bibnamefont{Moinester}} \emph{et~al.}
  (\bibinfo {collaboration} {SELEX}),\ }%
  \bibfield{journal}{%
  \bibinfo {journal} {Czech.J.Phys.}\ }%
  \textbf{\bibinfo {volume} {53}},\ \bibinfo {pages} {B201} (\bibinfo {year}
  {2003}),\ \Eprint{http://arxiv.org/abs/hep-ex/0212029}{arXiv:hep-ex/0212029
  [hep-ex]}%
  \bibAnnoteFile{NoStop}{Moinester:2002uw}%
\bibitem{Ocherashvili:2004hi}%
  \BibitemOpen
  \bibfield{author}{%
  \bibinfo {author} {\bibfnamefont{A.}~\bibnamefont{Ocherashvili}}
  \emph{et~al.} (\bibinfo {collaboration} {SELEX Collaboration}),\ }%
  \bibfield{journal}{%
  \Doi{10.1016/j.physletb.2005.09.043}{\bibinfo {journal} {Phys.Lett.}}\ }%
  \textbf{\bibinfo {volume} {B628}},\ \bibinfo {pages} {18} (\bibinfo {year}
  {2005}),\ \Eprint{http://arxiv.org/abs/hep-ex/0406033}{arXiv:hep-ex/0406033
  [hep-ex]}%
  \bibAnnoteFile{NoStop}{Ocherashvili:2004hi}%
\bibitem{Brodsky:2011zs}%
  \BibitemOpen
  \bibfield{author}{%
  \bibinfo {author} {\bibfnamefont{S.~J.}\ \bibnamefont{Brodsky}}, \bibinfo
  {author} {\bibfnamefont{F.-K.}\ \bibnamefont{Guo}}, \bibinfo {author}
  {\bibfnamefont{C.}~\bibnamefont{Hanhart}},\ and\ \bibinfo {author}
  {\bibfnamefont{U.-G.}\ \bibnamefont{Mei{\ss}ner}},\ }%
  \bibfield{journal}{%
  \Doi{10.1016/j.physletb.2011.03.014}{\bibinfo {journal} {Phys.Lett.}}\ }%
  \textbf{\bibinfo {volume} {B698}},\ \bibinfo {pages} {251} (\bibinfo {year}
  {2011}),\ \Eprint{http://arxiv.org/abs/1101.1983}{arXiv:1101.1983 [hep-ph]}%
  \bibAnnoteFile{NoStop}{Brodsky:2011zs}%
\bibitem{Namekawa:2013vu}%
  \BibitemOpen
  \bibfield{author}{%
  \bibinfo {author} {\bibfnamefont{Y.}~\bibnamefont{Namekawa}} \emph{et~al.}
  (\bibinfo {collaboration} {PACS-CS Collaboration})}%
   (\bibinfo {year} {2013}),\
  \Eprint{http://arxiv.org/abs/1301.4743}{arXiv:1301.4743 [hep-lat]}%
  \bibAnnoteFile{NoStop}{Namekawa:2013vu}%
\bibitem{Lewis:2008fu}%
  \BibitemOpen
  \bibfield{author}{%
  \bibinfo {author} {\bibfnamefont{R.}~\bibnamefont{Lewis}}\ and\ \bibinfo
  {author} {\bibfnamefont{R.}~\bibnamefont{Woloshyn}},\ }%
  \bibfield{journal}{%
  \Doi{10.1103/PhysRevD.79.014502}{\bibinfo {journal} {Phys.Rev.}}\ }%
  \textbf{\bibinfo {volume} {D79}},\ \bibinfo {pages} {014502} (\bibinfo {year}
  {2009}),\ \Eprint{http://arxiv.org/abs/0806.4783}{arXiv:0806.4783 [hep-lat]}%
  \bibAnnoteFile{NoStop}{Lewis:2008fu}%
\bibitem{Albertus:2009ww}%
  \BibitemOpen
  \bibfield{author}{%
  \bibinfo {author} {\bibfnamefont{C.}~\bibnamefont{Albertus}}, \bibinfo
  {author} {\bibfnamefont{E.}~\bibnamefont{Hernandez}},\ and\ \bibinfo {author}
  {\bibfnamefont{J.}~\bibnamefont{Nieves}},\ }%
  \bibfield{journal}{%
  \Doi{10.1016/j.physletb.2009.11.048}{\bibinfo {journal} {Phys.Lett.}}\ }%
  \textbf{\bibinfo {volume} {B683}},\ \bibinfo {pages} {21} (\bibinfo {year}
  {2010}),\ \Eprint{http://arxiv.org/abs/0911.0889}{arXiv:0911.0889 [hep-ph]}%
  \bibAnnoteFile{NoStop}{Albertus:2009ww}%
\bibitem{Genovese:1997tm}%
  \BibitemOpen
  \bibfield{author}{%
  \bibinfo {author} {\bibfnamefont{M.}~\bibnamefont{Genovese}}, \bibinfo
  {author} {\bibfnamefont{J.}~\bibnamefont{Richard}}, \bibinfo {author}
  {\bibfnamefont{F.}~\bibnamefont{Stancu}},\ and\ \bibinfo {author}
  {\bibfnamefont{S.}~\bibnamefont{Pepin}},\ }%
  \bibfield{journal}{%
  \Doi{10.1016/S0370-2693(98)00187-7}{\bibinfo {journal} {Phys.Lett.}}\ }%
  \textbf{\bibinfo {volume} {B425}},\ \bibinfo {pages} {171} (\bibinfo {year}
  {1998}),\ \Eprint{http://arxiv.org/abs/hep-ph/9712452}{arXiv:hep-ph/9712452
  [hep-ph]}%
  \bibAnnoteFile{NoStop}{Genovese:1997tm}%
\bibitem{Stewart:2004pd}%
  \BibitemOpen
  \bibfield{author}{%
  \bibinfo {author} {\bibfnamefont{I.~W.}\ \bibnamefont{Stewart}}, \bibinfo
  {author} {\bibfnamefont{M.~E.}\ \bibnamefont{Wessling}},\ and\ \bibinfo
  {author} {\bibfnamefont{M.~B.}\ \bibnamefont{Wise}},\ }%
  \bibfield{journal}{%
  \Doi{10.1016/j.physletb.2004.03.087}{\bibinfo {journal} {Phys.Lett.}}\ }%
  \textbf{\bibinfo {volume} {B590}},\ \bibinfo {pages} {185} (\bibinfo {year}
  {2004}),\ \Eprint{http://arxiv.org/abs/hep-ph/0402076}{arXiv:hep-ph/0402076
  [hep-ph]}%
  \bibAnnoteFile{NoStop}{Stewart:2004pd}%
\bibitem{Cohen:2005bx}%
  \BibitemOpen
  \bibfield{author}{%
  \bibinfo {author} {\bibfnamefont{T.~D.}\ \bibnamefont{Cohen}}, \bibinfo
  {author} {\bibfnamefont{P.~M.}\ \bibnamefont{Hohler}},\ and\ \bibinfo
  {author} {\bibfnamefont{R.~F.}\ \bibnamefont{Lebed}},\ }%
  \bibfield{journal}{%
  \Doi{10.1103/PhysRevD.72.074010}{\bibinfo {journal} {Phys.Rev.}}\ }%
  \textbf{\bibinfo {volume} {D72}},\ \bibinfo {pages} {074010} (\bibinfo {year}
  {2005}),\ \Eprint{http://arxiv.org/abs/hep-ph/0508199}{arXiv:hep-ph/0508199
  [hep-ph]}%
  \bibAnnoteFile{NoStop}{Cohen:2005bx}%

\end{thebibliography}%

\end{document}